\documentclass[aps,pra,showpacs,twocolumn,groupedaddress]{revtex4}
\usepackage{graphicx}
\usepackage{CJK}
\usepackage{amsmath}
\usepackage{amssymb}
\usepackage{subfigure}

\begin{document}
\draft
\title{Atom-molecule conversion with particle losses}
\author{B. Cui, L. C. Wang,  X. X. Yi}
\affiliation{School of Physics and Optoelectronic Technology,\\
Dalian University of Technology, Dalian 116024, China}

\date{\today}

\begin{abstract}
Based on the mean-field approximation and the phase space analysis,
we study the dynamics of an atom-molecule conversion system subject
to particle loss. Starting from the many-body dynamics described by
a master equation, an effective nonlinear Schr\"{o}dinger equation
is introduced. The classical phase space is then specified and
classified by  fixed points. The boundary, which separate different
dynamical regimes have been calculated and discussed. The effect of
particle loss on the conversion efficiency and the self-trapping is
explored.
\end{abstract}

 \maketitle

\section{introduction}
Association of ultracold atoms into molecules is currently an active
topic in the field of ultracold quantum physics, it attracts much
attention due to its  important applications ranging from the
production of molecular Bose-Einstein condensates to the search for
the permanent electric dipole moment, see for example
\cite{Heinzen00,Donley02,Hudson02,Greiner03,
Jochim03,Zwierlein03,Hines03,Regal04,Naidon08,Junker08,Jing09,Qian10}.
By applying a time varying magnetic field in the vicinity of
Feshbach resonance, a pair of atoms can bound into a diatomic
molecule \cite{Timmermans99,Kohler06}, this conversion can be
described by the Gross-Pitaevski (GP) equations within the
mean-field theory (MFT)
\cite{Vardi01,Santos06,Li09,Liu10,Fu10,Santos10}. Such an treatment
reduces the full many-body problem into a set of coupled nonlinear
Schr\"odinger equations and the complicated many-body dynamics is
then turned into a two-mode dynamics. Earlier study shows that the
nonlinearity, which arises from both the atom-atom and
molecule-molecule couplings, plays an important role in  the system.
Four distinct regimes, each has different feature in dynamics can be
classified, accordingly the bifurcation of the fixed points in the
classical phase space \cite{Santos06,Santos10} is identified.

Decoherence that arises from  the unavailable coupling between the
environment (thermal atoms or molecules)  and the condensed system
plays an important role in atomic or molecular Bose-Einstein
condensate
\cite{Anglin97,Ruostekoski98,Vardi011,Anglin01,Wang07,Syassen08}.
Description of  decoherence by fully including the quantum effects
requires sophisticated theoretical studies, however the standard
approach in  quantum optics can reduce the complexity and in fact it
has been widely used in   Bose-Einstein condensates
\cite{Trimborn08,Witthaut08,Witthaut09,Shchesnovich10,Graefe08,Graefe10}.
For an atom-molecule conversion system, we then ask:   will the
decoherence  effect be different from that in the pure atomic or
molecular Bose-Einstein condensates? What are the fixed points in
this situation? How do these fixed points behave? We will answer
these questions in this paper.

In this paper,  we focus on the amplitude decoherence (particle
loss), which may arise from inelastic collision between condensate
and noncondensate particles in the system. The standard approach in
quantum optics for open systems are used and the master equation is
derived by treating the noncondensate atoms and molecules as a
Markovian reservoir. Under the mean-field approximation, an
effective non-Hermitian Gross-Pitaevskii  equation is derived.
Bifurcation of the fixed points divides  the parameter space into
different dynamical regimes,  boundaries that separate  these
regimes are changed by the decoherence. By calculating the
corresponding Jacobian matrix, we find that a sudden transition in
the fixed point from elliptic point  to attractor or repeller
happens with non-zero  decoherence rate, which reflects the
meta-stable features of the system under the decoherence. The
atom-molecule conversion efficiency for the molecular condensate as
well as the self-trapping are also studied.

The paper is organized as follows. In Sec. \ref{md}, we introduce
the model and transform the master equation to a non-Hermitian
nonlinear Schr\"odinger equation. Conditions that determine the
fixed points are derived. In Sec. \ref{d1}, we define different
regimes by the bifurcation of the fixed points and study the
dynamics of system in these regimes. In Sec. \ref{d2}, we
investigate the effect of particle loss on the conversion
efficiency. In Sec. \ref{d3}, we shed light on the self-trapping
taking  the decoherence into account,  an explanation for the
observed features is given in the framework of mean-field theory.
Finally, we conclude our results in Sec. \ref{con}.

\section{model}\label{md}
Based on the two-mode approximation, the Hamiltonian that includes
the atom-atom collision $U_{aa}$, atom-molecule conversion with rate
$V$, and molecule-molecule couplings $U_{bb}$ takes the following
form \cite{Santos06,Li09}
\begin{eqnarray}
H&=&\mu_a\hat{{a}}^{\dag}\hat{a}
+\mu_b\hat{{b}}^{\dag}\hat{b}
+U_{aa}\hat{{a}}^{\dag}\hat{{a}}^{\dag}
\hat{a}\hat{a}+U_{bb}\hat{{b}}^{\dag}\hat{{b}}^{\dag}\hat{b}\hat{b}\\
&&+U_{ab}\hat{{a}}^{\dag}\hat{a}
\hat{{b}}^{\dag}\hat{b}+V(\hat{a}^{\dag}\hat{{a}}^{\dag}\hat{b}
+\hat{{b}}^{\dag}\hat{a}\hat{a}). \nonumber
\end{eqnarray}
 The master
equation \cite{Gardiner} that takes the  particle loss into account
can be  derived  as in the textbook \cite{Anglin97},
\begin{eqnarray}\label{mequation}
\dot{\rho}=&-&i[\hat{H},\rho]-\frac{\Gamma_a}{2}
(\hat{a}^{\dag}\hat{a}\rho+\rho\hat{{a}}^{\dag}\hat{a}-2\hat{a}\rho\hat{{a}}^{\dag})\\
&-&\frac{\Gamma_b}{2}(\hat{b}^{\dag}\hat{b}
\rho+\rho\hat{{b}}^{\dag}\hat{b}-2\hat{b}\rho\hat{{b}}^{\dag}),\nonumber
\end{eqnarray}
where $\Gamma_a$ and $\Gamma_b$ represent decoherence rates for
atomic and molecular modes, respectively. In the mean-field
approximation, the quantum fluctuation is negligible. It is
appropriate to replace $\hat{a}$ and $\hat{b}$ with $c$ numbers
$a=|a|e^{i\theta_a}$ and $b=|b|e^{i\theta_b}$. With these
considerations, the master equation (\ref{mequation}) can be casted
into the following nonlinear Schr\"{o}dinger equation,
\begin{eqnarray}\label{schrodinger}
i\frac{d}{dt}\begin{pmatrix}a\\b\end{pmatrix}=
H\begin{pmatrix}a\\b\end{pmatrix},
\end{eqnarray}
\begin{eqnarray}
H=\begin{pmatrix}R-U{z}-\frac{i}{2}
\Gamma_a&2Va^{*}\\Va&-2R+2Uz-\frac{i}{2}\Gamma_b\end{pmatrix},\label{hamilton}
\end{eqnarray}
with $z=|a|^2-2|b|^2$ describing  the  number  difference for atoms
in the two  modes.
$U=\frac{1}{4}U_{ab}-\frac{1}{2}U_{aa}-\frac{1}{8}U_{bb}$ represents
a coupling strength and $V$ is the conversion rate.
$R=\frac{1}{4}(2\mu_a-\mu_b+2U_{aa}-\frac{1}{2}U_{bb})$ denotes the
energy difference between the two modes, which can be effectively
tuned by a time-varying external field \cite{Li09,Liu10}. Here and
hereafter, we rescale $R$, $U$, $\Gamma_a$, $\Gamma_b$ in units of
$V$, and $t$ in units of $1/V$, $\hbar=1$ has been set, hence all
parameters in this paper are of dimensionless.

Similar to the two-mode Bose-Hubbard model, the projective Hilbert
space for such an atom-diatomic molecule conversion system can be
spanned by  a set of Bloch vectors. Under the mean-field
approximation, the Bloch vectors can be written as \cite{Fu10}
\begin{equation}
\textbf{h}=(2\sqrt{2}Re[(a^{*})^2b],2\sqrt{2}Im[(a^{*})^2b],|a|^2-2|b|^2).
\end{equation}
With the normalization condition $|a|^2+2|b|^2=1$,
the projective space is a tear-drop shaped surface as shown in Fig. \ref{blochs}.

To analyze the dissipative dynamics of the system in its classical phase space,
we define relative phase $\theta$, particle number $n$ and normalized
population difference $S$ as
\begin{eqnarray}
\theta&=&2\theta_a-\theta_b,\\
n&=&2|b|^2+|a|^2,\\
S&=&\frac{z}{n}.
\end{eqnarray}
Inserting these definitions into Eq. (\ref{schrodinger}), a set of
evolution equations is obtained
\begin{eqnarray}
\dot{S}&=&-2V\sqrt{n}(1+S)\sqrt{1-S}\sin{\theta}-\Gamma_{-}(1-S^2),\\
\dot{\theta}&=&4UnS-4R-V\sqrt{n}\frac{1-3S}{\sqrt{1-S}}\cos{\theta},\\
\dot{n}&=&-(\Gamma_{+}+\Gamma{-}{S})n,
\end{eqnarray}
with $\Gamma_{+}=\frac{1}{2}(\Gamma_a+\Gamma_b)$ and
$\Gamma_{-}=\frac{1}{2}(\Gamma_a-\Gamma_b)$ representing the total
and relative decoherence rates for  the two modes, respectively.
Ignoring the coupling between the system and its environment
($\Gamma_a=\Gamma_b=0$), the dynamics of the system can be described
by an effective Hamiltonian as
\begin{eqnarray}
H=2V(1+S)\sqrt{1-S}\cos{\theta}-2US^2+4RS.
\end{eqnarray}
It is found that the bifurcation of the fixed points falls into four
regimes in  the parameter space  \cite{Santos06,Santos10} as shown
in Fig. \ref{rgms}. Due to inelastic collisions between particles in
the condensates and that in the thermal cloud, the loss of atoms and
molecules as one source  of decoherence is unavoidable  in
practices. Thus it is interesting and desired to study the dynamics
of this system with decoherence. By  the standard procedure used in
non-Hermitian system \cite{Trimborn08,Graefe08}, we study the
dynamics of the system with notations  $C=Un$, $\Omega=V\sqrt{n}$.
With these knowledge, the effects of the particle loss on the
dynamics as well as the features of  fixed point can be clearly
revealed  in classical phase space. For simplicity,  in the
following discussion, we set $\Gamma=\Gamma_-$ and rewrite the
equations for the population difference and relative phase as
\begin{eqnarray}
\dot{S}&=&-2\Omega(1+S)\sqrt{1-S}\sin{\theta}-\Gamma(1-S^2),\label{pd}\\
\dot{\theta}&=&4CS-4R-\Omega\frac{1-3S}{\sqrt{1-S}}\cos{\theta}.\label{rp}
\end{eqnarray}
The fixed points of the system are determined by
\begin{equation}
\dot{S}=\dot{\theta}=0.
\end{equation}
To be specific, the fixed points on the boundary are  $S=-1$,
$\theta=\arccos({-\frac{\sqrt{2}(C+R)}{\Omega}})$, while the other
fixed points are determined by
\begin{eqnarray}
&&(9\Gamma^2+64C^2)S^3-(\Gamma^2-4\Omega^2+64R^2)\nonumber\\
&&-(15\Gamma^2-36\Omega^2+64C^2+128CR)S^2\nonumber\\
&&-(24\Omega^2-7\Gamma^2-64R^2-128CR)S=0,\label{podi}
\end{eqnarray}
\begin{eqnarray}
\sin{\theta}=-\frac{\Gamma}{2\Omega}\sqrt{1-S}.\label{reph}
\end{eqnarray}
Through a Jacobian matrix defined by
\begin{equation}
J=\left(
\begin{array}{ccc}
 {\partial{\dot{S}}}/{\partial{S}} &&
 {\partial{\dot{S}}}/{\partial{\theta}} \\
 {\partial{\dot{\theta}}}/{\partial{S}} &&
 {\partial{\dot{\theta}}}/{\partial{\theta}} \\
\end{array}
\right),
\label{jacobian}
\end{equation}
we can study  the stability of the fixed points as did in the
literature \cite{Graefe10,strogatz94,boyce97}.

From Eqs. (\ref{pd}) and ({\ref{rp}), we find that the dynamics of
the system  depends only on the relative decoherence  rate $\Gamma$,
but not on  the total decoherence rate. The total decoherence rate
affect the system through $n=n_0e^{-\Gamma_+t}$. With zero relative
decoherence rate ($\Gamma=0$), the effective Hamiltonian
can be written as \\
\begin{eqnarray}
H=2\Omega(1+S)\sqrt{1-S}\cos{\theta}-2CS^2+4RS,
\end{eqnarray}
which differs from the case without decoherence  only at the
conversion amplitude $\Omega$ and coupling strength $C$. In the next
section, we shall discuss  the effect of relative decoherence rate
on  the dynamics of the system.

\section{four dynamical regimes with  decoherence}\label{d1}
\begin{figure}
\includegraphics*[width=1.0\columnwidth,height=0.7\columnwidth]{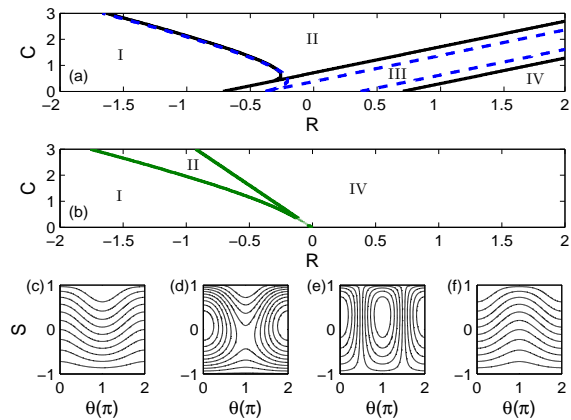}
\caption{(color online) Parameter space spanned  by nonlinearity $C$
and energy difference $R$. Different regimes are distinguished by
boundary lines, where black solid lines represent the case for
$\Gamma=0$, while  blue dashed lines denotes the case for
$\Gamma=1.2$ in (a). Green solid lines in (b) denotes the boundary
lines for $\Gamma=2.4$.  (c), (d), (e), and (f) describe the
classical phase space for region
$\uppercase\expandafter{\romannumeral1}$,
$\uppercase\expandafter{\romannumeral2}$,
$\uppercase\expandafter{\romannumeral3}$, and
$\uppercase\expandafter{\romannumeral4}$, respectively.}\label{rgms}
\end{figure}
\begin{figure}
\includegraphics*[width=0.8\columnwidth,height=0.9\columnwidth]{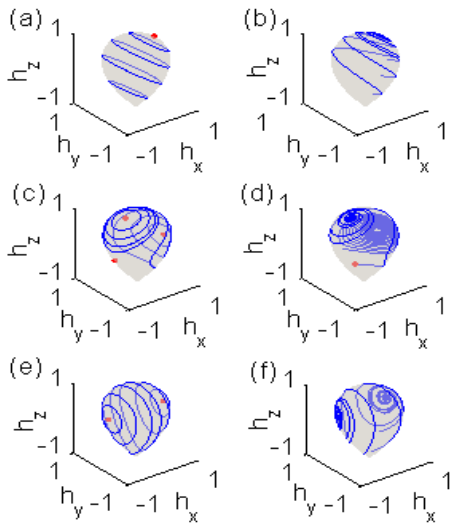}
\caption{(color online) Mean-field dynamics on the Bloch sphere for
Hermitian (left) and non-Hermitian (right) cases.  The north pole
and south pole of the sphere corresponds to the pure atomic
condensate and the pure molecular condensate, respectively. Red
spots and center of the vortex denote the location of the fixed
points. Blue solid lines repersent the trajectories for the time
evolution of the system. Parameters chosen are $R=1$, $U=0$ for (a)
and (b), $R=0$, $U=2$ for (c) and (d), and $R=0$, $U=0$ for (e) and
(f). The spheres on left side ((a),(c),(e)) describe decoherence
free case ($\Gamma=0$), while the spheres on right side depict the
decoherence case ($\Gamma=1$).}\label{blochs}
\end{figure}
\begin{figure}
\includegraphics*[width=0.9\columnwidth,height=0.5\columnwidth]{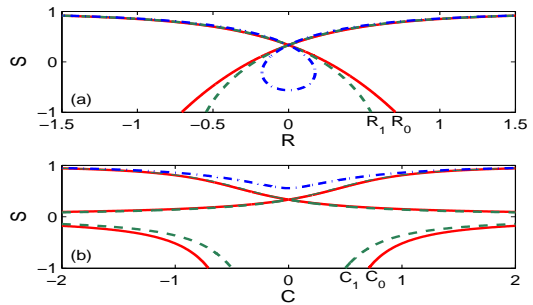}
\caption{(color online) (a) Locations of fixed points for different
decoherence rate $\Gamma$ and energy difference $R$. Parameters
chosen are $U=0$, $V=1$, and $\Gamma=0,0.9,1.6$ for red solid line,
green dashed line and blue dotted line, respectively. (b) Locations
of fixed points for different decoherence rate $\Gamma$ and
interaction strength $C$. Parameters chosen are $R=0$, $V=1$, and
$\Gamma=0,0.5,1.5$ for red solid line, green dashed line and blue
dotted line, respectively.}\label{fpn}
\end{figure}
In Ref. \cite{Santos06}, without considering the decoherence effect,
by the feature of  fixed points, the parameter space was divided
into  four regions. Here we re-divide the region by taking the
decoherence into account (see Fig. \ref{rgms} and Fig.
\ref{blochs}). Boundaries that  separate  different regions, are
determined by numerically solving Eqs. (\ref{podi},\ref{reph}). Note
that the fixed points on the boundary behave like the fixed points
in the region labeled by a  smaller number (e.g. boundary that
separate  regions $\uppercase\expandafter{\romannumeral1}$ and
$\uppercase\expandafter{\romannumeral2}$ belongs to  the region
$\uppercase\expandafter{\romannumeral1}$).

Figure \ref{rgms}(c) shows Poincar\'e section of the classical
Hamiltonian  for the region
$\uppercase\expandafter{\romannumeral1}$. The only fixed point is
located near the boundary of the phase space ($S=1$) and the
dynamics of the system is localized. When taking the decoherence
into consideration, the fixed point near $S=1$ turns into an
attractor, where Figs. \ref{blochs}(a) and \ref{blochs}(b) show
trajectories  on the tear-drop shaped Bloch sphere. The dynamics of
the system becomes delocalized due to the appearance of  such an
attractor.

By changing  the energy difference $R$ and the nonlinearity $C$ (see
Fig. \ref{rgms}(a)), the system can go across the boundary into
region $\uppercase\expandafter{\romannumeral2},$ the fixed point in
 region $\uppercase\expandafter{\romannumeral1}$ bifurcates into
two elliptic points and a hyperbolic one as Fig. \ref{rgms}(d)
shows. The region $\uppercase\expandafter{\romannumeral2}$ shares
similar features  with the self-trapping  in the two-mode
Bose-Hubburd model \cite{smerzi97,albiez05}. With  a negative
decoherence rate, both of the two elliptic fixed points transit to
attractors in this region (see Figs. \ref{blochs}(c) and
\ref{blochs}(d)). While the locations of the stable attractors are
just slightly changed due to the decoherence (see Fig.
\ref{fpn}(b)).

Figure \ref{rgms}(e) illustrates the Poincar\'{e} section of the
classical Hamiltonian  for region
$\uppercase\expandafter{\romannumeral3}$ without decoherence. In
this region, large amplitude oscillations around the elliptic fixed
can be observed, see Fig. \ref{blochs}(e). With  $U=0$ and $R=0$,
the location of the fixed points in this region can be derived
analytically
\begin{eqnarray}
(S,\theta)=\begin{pmatrix}
\frac{1}{3},&&\pi+\arcsin{(\frac{\Gamma}{\sqrt{6}\Omega})}\\
\frac{1}{3},&&2\pi-\arcsin{(\frac{\Gamma}{\sqrt{6}\Omega})}\end{pmatrix},\label{stheta}
\end{eqnarray}
where we assume the decoherence  rate  positive,  and the relative
phase was restricted  in $\theta\in[0,2\pi]$. From Eq.
(\ref{stheta}), we find that in addition to the feature changes of
 the fixed points, the relative phase between
the two fixed points  decreases  and the fixed points becomes
asymmetric due to the  decoherence as shown in Fig. \ref{blochs}(f).
As the decoherence rate  increases, the area of  regime
$\uppercase\expandafter{\romannumeral3}$ is compressed (see blue
dashed line in Fig. \ref{rgms}(a)). The two boundaries  coincides
and region $\uppercase\expandafter{\romannumeral3}$ vanishes (see
dash dotted line in Fig. \ref{fpn}(a)), when decoherence rate is
larger than a threshold ($\Gamma>\sqrt{2}\Omega$), a hyperbolic
fixed point arises  from the bottom of the phase space (see
dash-dotted line in Fig. \ref{fpn}(a)). The boundary that separates
regions  $\uppercase\expandafter{\romannumeral3}$ and
$\uppercase\expandafter{\romannumeral4}$ is shifted due to
decoherence. This boundary shift can be explained as  a threshold
decrease in the energy difference $R$ (denoted by $R_0$ and $R_1$ in
Fig. \ref{fpn}(a)), which is an witness for the bifurcation of fixed
points in classical phase space.

The dynamics in region $\uppercase\expandafter{\romannumeral4}$
behaves similarly as that in  region
$\uppercase\expandafter{\romannumeral1}$. The elliptic fixed point
turns into an attractor due to  negative decoherence rate,  the
dynamics in this region then becomes delocalized (see Figs.
\ref{blochs}(a) and \ref{blochs}(b)).

Next, we focus on the  changes of the fixed points. Such a change in
classical phase space is fundamental for non-hermitian Bose-Hubbard
system \cite{Trimborn08,Graefe08,Graefe10}. However, we find that,
in the atom-molecule conversion system, the change differs from
Bose-Hubbard model in two respects. Firstly, the type of the fixed
point (a repeller or an attractor) is determined by the the sign of
decoherence rate $\Gamma$ and the location of the fixed point $S$.
If $\Gamma$ and $S$  are different in sign, i.e., one of them is
positive while another is negative,  the original elliptic fixed
point transits into a stable attractor. Otherwise, the original
fixed point turns into an unstable repeller. Secondly, the
transition is {\it sudden}, in other words, the transition happens
provided the decoherence rate is not zero. This is different from
the decoherence effect on Bose-Einstein condensates in a double-well
potential, namely there exists a critical value for  the decoherence
rate \cite{Trimborn08}. In the atom-molecule conversion  system, the
transition happens once the decoherence exists, regardless of how
small the decoherence being. The phenomenon reflects not only the
meta-stable behavior of the open many-particle system, but also the
sensitivity of the atom-molecule conversion system to particle loss.

\section{conversion efficiency for molecular condensate}\label{d2}

\begin{figure}
\includegraphics*[width=0.9\columnwidth,height=0.5\columnwidth]{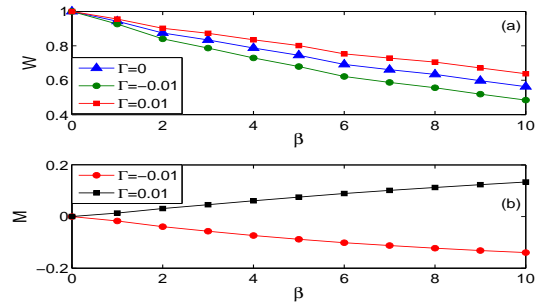}
\caption{(color online) Conversion efficiency $W$ in (a) and
relative efficiency $M$ in (b) as a function of the sweeping rate
$\beta$ under different decoherence rate $\Gamma$.}\label{eff}
\end{figure}
In experiments, the association of ultracold atoms into diatomic
molecules can be achieved by applying a time-dependent magnetic
field in the vicinity of a Feshbach resonance, which corresponds to
the change between different regimes
($\uppercase\expandafter{\romannumeral1}\rightarrow
\uppercase\expandafter{\romannumeral3}\rightarrow\uppercase\expandafter{\romannumeral4}$)
in the parameter space (see Fig. \ref{rgms}). To examine the effect
of decoherence on the conversion process, we define conversion
efficiency, relative efficiency and sweeping rate of the external
field as
\begin{eqnarray}
W&=&\frac{|b(T)|^2}{n(T)},\\
M&=&\frac{W(\Gamma)-W(0)}{W(0)},\\
\beta&=&\dot{R},
\end{eqnarray}
where $T$ denotes the terminal time for the conversion, $W(\Gamma)$
and $W(0)$ denote the conversion efficiency with and without
decoherence, respectively. $M$ describes the relative increases  or
decreases  of the efficiency between decoherence and
decoherence-free case. By adjusting the external magnetic field
\cite{Li09}, $R$ can be linearly manipulated  to across the Feshbach
resonance point($R=R_0+\beta{t}, R_0=\beta{T}, t\in[0,T]$), until
the system relaxes into a steady state. The conversion efficiency
with decoherence can be  calculated with  the same parameters except
the decoherence rate. Here we choose the initial state of the system
 to be in  pure atomic mode ($|a(0)|^2=1$).

The results of  $W$ show that conversion efficiency  increases  with
positive decoherence rate. While a negative decoherence rate
decreases  the  conversion efficiency (see Fig. \ref{eff}). This can
be interpreted by the appearance  of attractor or repeller in the
phase space. With  a negative decoherence rate, the elliptic fixed
points near the atomic mode turns into an attractor and the atoms
are attracted to stay away from molecular mode. (see Figs.
\ref{blochs}(b) and \ref{blochs}(f)). The conversion process is
depressed  by such an attractor and the conversion efficiency
decreases. Similarly,  a positive decoherence rate will increase the
conversion efficiency.

\section{tunneling and self-trapping}\label{d3}

\begin{figure}
\includegraphics*[width=0.9\columnwidth,height=0.5\columnwidth]{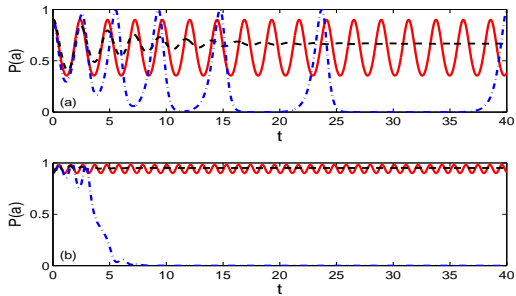}
\caption{(color online) Time evolution for the population of atomic
mode $P(a)=|a(t)|^2$ under different decoherence rates as
$\Gamma=0,-0.5,0.5$ denoted by red solid line, black dashed line and
blue dash-dotted line both in (a) and (b). Parameters chosen are
$V=1$, $R=0$ for both (a) and (b), $U=0$ for (a) and $U=1.5$ for
(b). The initial population for atoms are $|a(0)|^2=0.9$.}\label{pt}
\end{figure}
In this section, we investigate the effect of particle loss on the
dynamics  of the system, the atoms may oscillates between atomic and
molecular modes (corresponding  to regime
\uppercase\expandafter{\romannumeral3}), and they can also be
trapped in one of the modes (corresponding to  the regime
\uppercase\expandafter{\romannumeral2} in parameter space).

In regime \uppercase\expandafter{\romannumeral3}, the atoms
oscillate  between atomic mode and molecular mode (see Fig.
\ref{blochs}(e)). When the relative decoherence rate is positive,
the fixed point transits from elliptic to a repeller, the amplitude
of the oscillation is then increased  (see dash dotted line in Fig.
\ref{pt}(a)).  While for negative relative decoherence rate, the
oscillation is compressed, since the elliptic fixed point suddenly
transits to an attractor (see dashed line in Fig. \ref{pt}(a)).

With $C$ increases,  the dynamics of the system turns into the
self-trapping regime, which belongs  to the regime
\uppercase\expandafter{\romannumeral2} in Fig. \ref{rgms}(a). We
find that the threshold of the coupling constant  is decreased by
the decoherence, i.e., the decoherence supports the self-trapping
(denoted by $C_0$ and $C_1$ in Fig. \ref{fpn}(b)). With  negative
relative  decoherence rate, the fixed point near the atomic mode
transits into an attractor. The self-trapping in atomic mode
keeps(see black dashed line in Fig. \ref{pt}(b)). When the relative
decoherence rate is positive, which indicates a repeller in the
phase space, the self-trapping in atomic mode is ruined,  because
the atoms are repelled and  converted into molecules, as dash dotted
line shows in Fig. \ref{pt}(b).

A physics understanding is coming .......

\section{conclusion}\label{con}
In summary, we have investigated  the effect of particle loss on the
dynamics of the atom-molecule conversion system. Within  the
mean-field approximation, the classical phase space is specified and
the fixed points are calculated. Due to the bifurcation of the fixe
points in the phase space, the parameter space can be divided  into
different regimes. We find that the boundary, which separates
different regimes are changed by the decoherence. A sudden
transition for the  fixed points from elliptic to attractor or
repeller happens. Such a transition not only reflects the
meta-stable behavior of the system,  but also characterizes the
phase-space structure of the atom-molecule conversion system.  The
effect of decoherence on the conversion efficiency and the
self-trapping is explored with the mean-field approximation.

This work is supported by NSF of China under grant Nos 61078011 and
10935010, the Open Research Fund of State Key Laboratory of
Precision Spectroscopy, East China Normal University, as well as the
National Research Foundation and Ministry of Education, Singapore
under academic research grant No. WBS: R-710-000-008-271.


\end{document}